\documentclass[aps,onecolumn,11pt,floatfix,altaffilletter,superscriptaddress,preprintnumbers, tightenlines,showpacs,showkeys,notitlepage,nofootinbib]{revtex4-2}

\usepackage[colorlinks=true,citecolor=blue,linkcolor=blue]{hyperref}
\usepackage[normalem]{ulem}
\usepackage{amsmath,amssymb}
\usepackage{mathtools}
\usepackage{verbatim}
\usepackage{titlesec}               
\usepackage{epsfig}
\usepackage{graphicx}               
\usepackage{url}
\usepackage{color}
\usepackage{multirow}
\usepackage{floatrow}
\usepackage{placeins}
\usepackage[dvipsnames]{xcolor}
\usepackage{epstopdf}
\usepackage{siunitx}
\usepackage{fontawesome}
\usepackage{tikz}
\usepackage{tikz-feynman}
\usepackage{enumitem}
\usepackage[capitalize]{cleveref}
\usepackage{lipsum}
\usepackage{gensymb}
\usepackage{booktabs}               
\usepackage{xspace}                 
\usepackage{pifont}
\usepackage{marvosym }
\usetikzlibrary{shapes,arrows}
\usetikzlibrary{decorations.pathmorphing,decorations.markings}
\usetikzlibrary{snakes}
\usepackage{orcidlink}  
\usepackage{fontawesome}
\newcommand{\gitlink}{\href{https://github.com/samiur06/cosmogenic_neutrino_LIV}{\textsc{g}it\textsc{h}ub~{\large\color{black}\faGithub}}\xspace}
\allowdisplaybreaks

\usepackage{diagbox}
\usepackage{tabularx}
\usepackage{array}

\usepackage{bbm}
\usepackage{slashed}

\makeatletter

\renewcommand{\p@subsection}{}
\makeatother

\titleformat*{\section}{\centering\bfseries\uppercase}
\titlelabel{\thetitle\quad}
\titleformat*{\paragraph}{\bfseries}
\titlespacing*{\paragraph}{0pt}{3.25ex plus 1ex minus .2ex}{1em}

\makeatletter
\def\l@subsubsection#1#2{}
\makeatother

\newsavebox{\twosubbox}

\begin{document}

\title{Ultra-High-Energy Tau Neutrinos as Probes of Lorentz Invariance}

\author{Vedran Brdar \orcidlink{0000-0001-7027-5104}}
\email{vedran.brdar@okstate.edu}
\affiliation{Department of Physics, Oklahoma State University, Stillwater, OK, 74078, USA}
\author{Samiur R. Mir \orcidlink{0000-0002-6531-2174} }
\email{samiur.mir@okstate.edu}
\affiliation{Department of Physics, Oklahoma State University, Stillwater, OK, 74078, USA}

\begin{abstract}
Neutrino telescopes have detected astrophysical neutrinos with energies up to
$\mathcal{O}(100)$ PeV. Several current and proposed experiments aim to observe
neutrinos at even higher energies, with the goal of detecting cosmogenic
neutrinos. This increase in neutrino energy makes tests of Lorentz invariance
violation (LIV) particularly appealing, since the effects of higher-dimension LIV
operators on neutrino propagation grow rapidly with energy.
In this work, we investigate the potential of the upcoming experiments GRAND and
POEMMA to probe LIV in the neutrino sector through the detection of ultra-high-energy
tau neutrinos. We generate the cosmogenic neutrino flux using \texttt{SimProp} and
interface it with a calculation of neutrino flavor transition probabilities in the presence of
LIV effects. Deviations from standard flavor transition probabilities 
manifest as changes in the expected tau neutrino event rates at GRAND and POEMMA. We first consider the case with a single nonzero LIV operator of various dimensions, and find that the projected sensitivities exceed existing limits from lower-energy probes by orders of magnitude. We then explore scenarios with multiple nonzero LIV parameters and show that their interplay can significantly modify the sensitivities compared to the single-parameter case. Overall, we find that upcoming observations of ultra-high-energy tau neutrinos will place some of the most stringent constraints on LIV.
\end{abstract}

\maketitle

\section{Introduction}
\label{sec:intro}
\noindent
In 2013, the IceCube Neutrino Observatory pioneered the detection of neutrinos with energies at the PeV scale \cite{IceCube:2013low, IceCube:2014stg}. Neutrinos from previously observed astrophysical sources, namely those from the Sun \cite{Davis:1968cp,Cleveland:1998nv,SAGE:1999nng,SNO:2002tuh,Super-Kamiokande:2001ljr,
Borexino:2007kvk} and the supernova 1987A \cite{Kamiokande-II:1987idp,Bionta:1987qt,Baksan}, have typical energies of $\mathcal{O}(10)$ MeV, meaning that the IceCube discovery increased the accessible astrophysical neutrino energy range by nine orders of magnitude. In the last decade, IceCube has been steadily accumulating high-energy neutrino events
\cite{IceCube:2018cha,IceCube:2020wum,IceCube:2024fxo} including the detection of Glashow resonance at $\sim 6.3$ PeV \cite{IceCube:2021rpz}. In 2025, the KM3NeT collaboration reported the observation of a $220$ PeV neutrino \cite{KM3NeT:2025npi,KM3NeT:2025vut,KM3NeT:2025bxl}, setting a new world record for the highest-energy neutrino detected. Note, however, that there is tension between the KM3NeT observation and the IceCube non-observation of $\mathcal{O}(100)$ PeV neutrinos \cite{Li:2025tqf} which can be resolved with production mechanisms involving new physics \cite{Brdar:2025azm,Dev:2025czz,Farzan:2025ydi}. 
Given the success of IceCube and KM3NeT, it is natural to ask: $(i)$ up to what energy scale neutrinos can be detected in the foreseeable future and $(ii)$ what scenarios can be tested with such observations?

Regarding the first question, with the further increase of neutrino energy toward the EeV scale and above, the likelihood of neutrino detection at the IceCube and KM3NeT experiments diminishes, as the astrophysical flux decreases with energy faster than the neutrino effective area grows. In addition, for neutrinos of such energy, Earth absorption becomes strong, further reducing the probability for neutrino detection. Nevertheless, there are several operating and proposed Earth- and space-based experiments with much larger effective area that are designed for the detection of extreme-energy neutrinos. These include Beacon \cite{Southall:2022yil}, GRAND \cite{GRAND:2018iaj}, IceCube-Gen2 radio \cite{IceCube-Gen2:2021rkf}, POEMMA \cite{POEMMA:2020ykm}, PUEO \cite{PUEO:2020bnn} and Trinity \cite{Otte:2025dld} (for review, see Ref.~\cite{Ackermann:2022rqc}) which share a common goal of detecting cosmogenic neutrinos \cite{Greisen:1966jv,Zatsepin:1966jv,Halzen:2002pg}. In this work, we will focus on GRAND and POEMMA, which are specialized for tau neutrino detection through radio and optical methods, respectively.

Regarding the second question, the scenarios that can be probed with ultra-high-energy neutrino data involve a wide range of neutrinophilic new physics models such as neutrino self-interactions \cite{Brdar:2022kpu,Leal:2025eou,Maitra:2025opp,Machado:2025ltu}, neutrino-dark matter interactions \cite{Leal:2025eou}, and neutrino transition magnetic moments \cite{Huang:2022pce}. However, scenarios that are particularly pronounced at ultra-high neutrino energies only arise
if fundamental symmetries such as Lorentz invariance are broken~\cite{Colladay:1998fq,Kostelecky:2008ts}.

Quantum gravity may induce the breaking of Lorentz symmetry around the Planck scale. At energies below the Planck scale, Lorentz invariance violation (LIV) can then be described within an effective field theory framework through a tower of higher-dimension operators suppressed by powers of the Planck mass. 
These operators, as can be verified through dimensional analysis, exhibit an extremely strong energy dependence. For the general formalism describing LIV and existing constraints, see Refs.~\cite{Colladay:1998fq,Kostelecky:2008ts}; for a dedicated discussion of the neutrino sector, we refer the reader to Ref.~\cite{Kostelecky:2003cr}.

LIV has been extensively explored using high-energy neutrino data from IceCube and KM3NeT \cite{Arguelles:2015dca,IceCube:2021tdn,Telalovic:2023tcb,Bustamante:2024fbj,Telalovic:2025xor,KM3NeT:2025mfl,Satunin:2025uui,Yang:2025kfr,Li:2025yvq}. In Ref.~\cite{Testagrossa:2023ukh}, the authors presented the sensitivity of GRAND to LIV effects by focusing on scenarios with a single nonzero LIV operator of various dimensions. In this work, we complement and extend these efforts. Namely, in addition to GRAND, we perform the analysis for POEMMA, and beyond the single-operator framework, we also explore scenarios with multiple nonzero LIV operators, whose interplay can significantly modify the sensitivities compared to cases where only a single LIV operator is present. Our approach to generating the cosmogenic neutrino flux is based on \texttt{SimProp} \cite{Aloisio:2012wj,Aloisio:2015sga,Aloisio:2016tqp,Aloisio:2017iyh}. \textcolor{black}{This leads to flux predictions that differ from those used in Ref.~\cite{Testagrossa:2023ukh} where the authors were instead motivated by the ultra-high-energy astrophysical neutrino flux from newborn pulsars \cite{Fang:2013vla}.}

The paper is organized as follows. In~\cref{sec:flavor_ratio_evolution}, we introduce the general formalism for LIV in the neutrino sector, with emphasis on the additional energy dependence in neutrino flavor evolution induced by LIV. We also show explicitly how LIV operators modify the tau neutrino fraction of cosmogenic neutrinos. 
In~\cref{sec:flux_det_count}, we present the details of the cosmogenic neutrino flux calculation and describe the analysis pipeline used to compute the number of tau neutrino events at GRAND and POEMMA in the presence of LIV. Our code for the cosmogenic neutrino flux treatment using \texttt{SimProp}, along with the LIV implementation, is available on~\gitlink. This framework allows us to derive projected sensitivities for the coefficients of LIV operators of different dimensions. Sensitivities for scenarios with both single and multiple nonzero LIV parameters are presented in~\cref{sec:exp_sens}. 
Our results show that experiments capable of detecting ultra-high-energy tau neutrinos have the potential to set leading constraints on the coefficients of LIV operators in the neutrino sector. We conclude in~\cref{sec:conclusions}.

\section{Neutrino Flavor Evolution in the presence of LIV}
\label{sec:flavor_ratio_evolution}
\noindent
The evolution of neutrino flavor states in the presence of LIV is governed by the following Hamiltonian
\begin{align}
    H_{\mathrm {tot}} = H_0 + H_{\mathrm {LIV}}\,. 
    \label{eq:H}
\end{align}
The first term in \cref{eq:H} is the vacuum Hamiltonian,
\begin{align}
    H_0 = U \,\rm{diag} \left(0,\frac{\Delta m_{21} ^2}{2E},\frac{\Delta m_{31} ^2}{2E}\right)\, U^ \dagger,
    \label{eq:H0}
\end{align}
where $U$ is the leptonic mixing matrix, $\Delta m^2_{ij}$ is the mass-squared difference of neutrino masses $m_i$ and $m_j$, and $E$ is the neutrino energy.

The second term in \cref{eq:H} encodes contributions from LIV operators affecting neutrino propagation \cite{Kostelecky:2011gq}. It reads
\begin{equation}
H_{\rm LIV} = \frac{1}{E}\left(\hat{a}_{\rm eff} - \hat{c}_{\rm eff}\right)\,,
\label{eq:HLIV_a-c}
\end{equation}
where $\hat{a}_{\rm eff}$ and $\hat{c}_{\rm eff}$ are matrices in flavor space corresponding to CPT-odd and CPT-even coefficients, respectively, arising from LIV operators of arbitrary dimension. One can rewrite $H_{\rm LIV}$ in terms of flavor-dependent LIV parameters as \cite{Kostelecky:2011gq,IceCube:2017qyp,Telalovic:2025xor}
\begin{align}
    H_{\rm LIV} = \sum_{d=2} ^\infty  (-1)^{d-1}{E} ^{d-3}  \begin{pmatrix}
\mathring{\kappa}^{(d)}_{ee} & \mathring{\kappa}^{(d)}_{e\mu} & \mathring{\kappa}^{(d)}_{e\tau} \\
\mathring{\kappa}^{(d)}_{\mu e} & \mathring{\kappa}^{(d)}_{\mu\mu} & \mathring{\kappa}^{(d)}_{\mu\tau} \\
\mathring{\kappa}^{(d)}_{\tau e} & \mathring{\kappa}^{(d)}_{\tau\mu} & \mathring{\kappa}^{(d)}_{\tau\tau}
\end{pmatrix}\,, 
\label{eq:HLIV_kappa}
\end{align}
where $\mathring{\kappa}^{(d)}_{\alpha \beta}$ is the isotropic LIV parameter corresponding to the flavor indices $\{\alpha,\beta\}$ and dimension $d$.  In our analysis, we consider only real values of $\mathring{\kappa}^{(d)}_{\alpha \beta}$. Since $H_{\rm LIV}$ is Hermitian, this implies $\mathring{\kappa}^{(d)}_{\alpha \beta} = \mathring{\kappa}^{(d)}_{\beta \alpha}$, leading to six independent LIV parameters at a given dimension $d$.

We note that the integer $d$ does not necessarily correspond to the mass dimension of the underlying LIV operators for massive neutrinos~\cite{Kostelecky:2011gq}. Since neutrino masses are negligible in our analysis, we refer to $d$ simply as the `dimension' in what follows. Because we are interested in neutrino interactions at ultra-high energies, we exclude $d=2$ scenario from our consideration, as in this case the LIV effect does not grow with energy $(H_{\rm LIV} \propto E^{-1})$.

In the presence of LIV, the modified leptonic mixing matrix $V(E)$ is obtained by diagonalizing $H_{\mathrm{tot}}$. In the limit where $H_{\rm LIV}$ is subdominant compared to the vacuum Hamiltonian ($H_{\rm LIV} \ll H_0$), $V(E)$ converges to the standard leptonic mixing matrix $U$. For a single nonzero LIV parameter $\mathring{\kappa}^{(d)}_{\alpha \beta}$, $H_{\rm LIV}$ becomes comparable to the vacuum Hamiltonian when
\begin{align}
    \mathring{\kappa}^{(d)}_{\alpha \beta}  \approx  10^{-2} \left(\frac{\Delta m_{ij} ^2}{E ^{d-2}} \right) =  10^{-23}\left( \frac{\Delta m_{ij} ^2}{10^{-3}\rm ~eV^2} \right) \left( \frac{\rm GeV}{E} \right)^{d-2} \text{GeV}^{4-d}\,.
\label{eq:kappa_compare_LIV_vac}
\end{align}
Here, the prefactor $10^{-2}$ represents an estimated contribution from the matrix elements of $U$ and the $1/2$ factor in~\cref{eq:H0}. \cref{eq:kappa_compare_LIV_vac} provides a rough estimate of the sensitivity expected for the parameters $\mathring{\kappa}^{(d)}_{\alpha \beta}$. Note the $E^{2-d}$ dependence in this equation; this shows that the parameters that can be probed become smaller as $d$ increases. 
This is especially pronounced for extreme-energy neutrinos. Stringent constraints on flavor-sensitive LIV parameters have already been reported using IceCube data in Refs.~\cite{IceCube:2021tdn,Telalovic:2025xor}. Ref.~\cite{IceCube:2021tdn} presents limits on isotropic LIV parameters, while Ref.~\cite{Telalovic:2025xor} provides an extensive analysis for direction-dependent LIV parameters over the range $d=2$ to $d=8$.

Let us now turn to the production and flavor evolution of cosmogenic neutrinos.\footnote{Since neutrinos and antineutrinos will produce indistinguishable signatures in the considered experiments, we will refer to both species as neutrinos for brevity.} 
Cosmogenic neutrinos are efficiently produced when high-energy \textcolor{black}{protons or nuclei} from ultra-high-energy cosmic rays (UHECRs) scatter off cosmic microwave background (CMB) or extragalactic background light (EBL), resulting in pion production \textcolor{black}{or radioactive beta decay of a neutron or unstable nucleus} \cite{Berezinsky:1969erk,Engel:2001hd}. Regarding the former, the charged pion decay ($\pi^+ \to \mu^+ \nu_\mu$) is followed by a muon decay ($\mu^+ \to e^+ \nu_e \bar{\nu}_\mu$); therefore, the neutrino flavor composition at the source is $(f_e : f_\mu : f_\tau)_S = (1/3:2/3:0)$~\cite{Beacom:2003nh,Athar:2005wg}. \textcolor{black}{As for the latter, beta decay yields electron neutrino in the final state. For such production, the flavor ratio therefore reads $(f_e : f_\mu : f_\tau)_S = (1:0:0)$ \cite{Lipari:2007su}.} Since the production of cosmogenic neutrinos occurs primarily in intergalactic space during UHECR propagation, in environments with small magnetic fields and low matter densities, there is no opportunity for muons from pion decay to lose energy before decaying, implying that $(f_e : f_\mu : f_\tau)_S = (0:1:0)$~\cite{Rachen:1998fd,Kashti:2005qa,Kachelriess:2006ksy,Lipari:2007su} is not realized. 

The flavor transition probability, for a neutrino $\nu_\alpha$ produced at redshift $z$ and detected as $\nu_\beta$ at $z=0$, reads
\begin{align}
    P_{\alpha\beta}(E)=\sum_{i=1}^{3} \left| V_{\alpha i}(E(1+z)) \right|^2 \left| V_{\beta i}(E) \right|^2\,,
    \label{eq:probb}
\end{align}
where $E$ is the neutrino energy at detection, and $E(1+z)$ corresponds to the neutrino energy at the production site.

One can also define the energy-dependent flavor fraction at Earth as
\begin{align}
    f_{\beta,\oplus}(E) = \sum_{\beta=e,\mu,\tau} P_{\alpha\beta}(E)\, f_{\alpha,S},
    \label{eq:ao}
\end{align}
where $f_{\alpha,S}$ is the normalized flavor composition of flavor $\alpha$ at the source. 

\begin{figure}[t]
  \centering
  \begin{tabular}{cc}
    \includegraphics[width=0.48\textwidth]{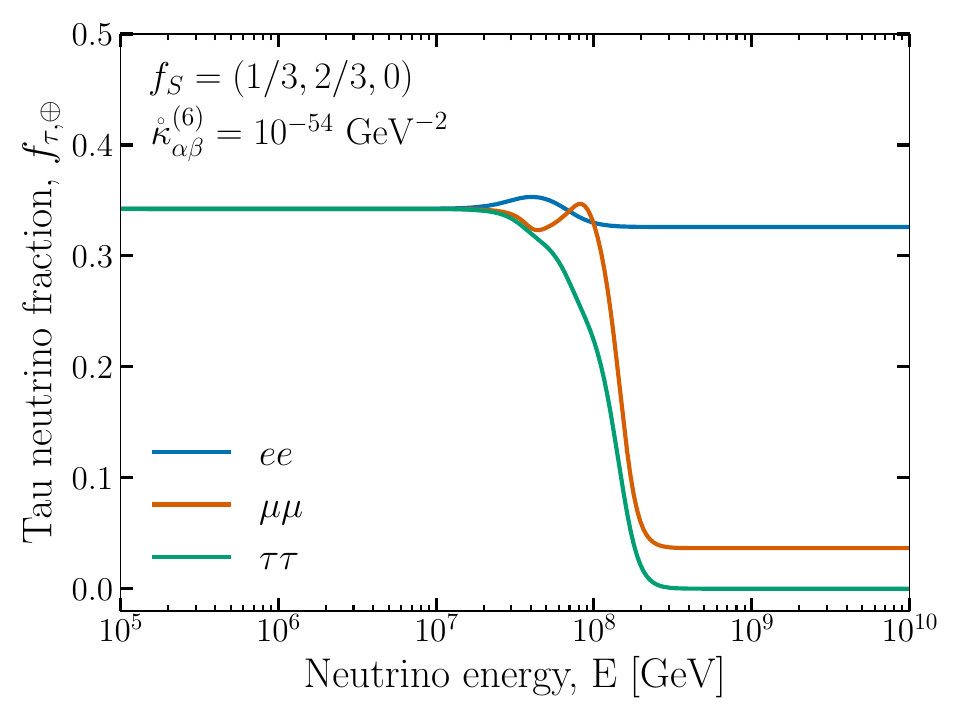}  &
    \includegraphics[width=0.48\textwidth]{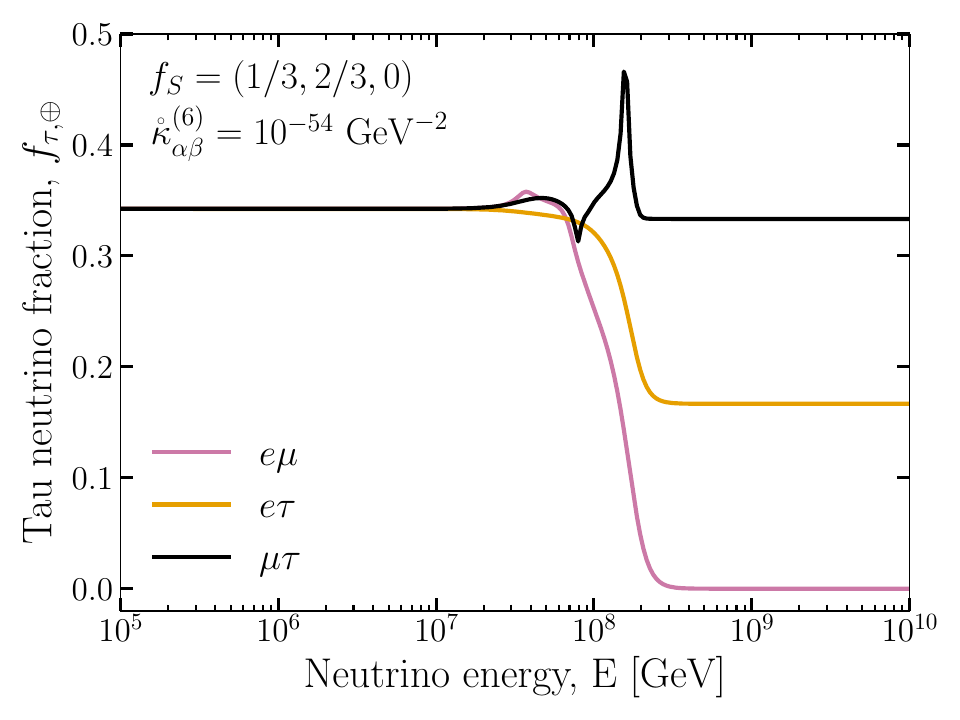}  \\ 
  \end{tabular}
    \caption{Tau neutrino flavor fraction at Earth, $f_{\tau,\oplus}$, as a function of the neutrino energy assuming neutrino production at redshift $z=1$. In the left (right) panel, a single nonzero dimension-6 LIV parameter corresponding to identical (mixed) flavor indices is considered.}
    \label{fig:flavor_fraction}
\end{figure}


In \cref{fig:flavor_fraction}, we show the tau neutrino fraction, $f_{\tau,\oplus}$, as a function of the neutrino energy at detection for a fixed value of a single nonzero dimension-6 LIV parameter, \textcolor{black}{assuming a source flavor composition of $(f_e : f_\mu : f_\tau)_S = (1/3:2/3:0)$}.
Regarding LIV parameters, we consider both scenarios corresponding to identical flavor indices ($\alpha=\beta$, left panel) and mixed flavor indices ($\alpha\neq\beta$, right panel). We fix the redshift at neutrino production to $z=1$. Generally, we observe three regions in the figure. At low energies, the LIV effect is not pronounced, and the flavor composition converges to $(f_e : f_\mu : f_\tau)_\oplus = (1/3:1/3:1/3)$. At very high energies, the LIV term dominates in $H_\text{tot}$, and the resulting composition is also constant, but $f_{\tau,\oplus}$ typically takes markedly different values, the exception being the $ee$ and $\mu\tau$ cases where the tau neutrino fraction attains values similar to those in the low-energy regime. Finally, there is an intermediate regime interpolating between the two regions discussed above.

The $\mu\tau$ case in \cref{fig:flavor_fraction} features a sharp peak around $E\sim 2\times 10^8$ GeV. We have explored this effect analytically in the two-flavor approximation and found that it is expected when
\begin{align}
    \mathring{\kappa}^{(d)}_{\alpha \neq \beta} = \frac{(-1)^d}{4} {\Delta m^2}\, E^{2 - d}\, \sin 2\theta \,.
    \label{eq:2f}
\end{align}
Here, $\theta$ is the mixing angle and $\Delta m^2$ is the mass-squared difference. Indeed, taking atmospheric oscillation parameters, i.e., $\Delta m^2= 2.5\times 10^{-3}\,\mathrm{eV}^2$, $\theta=\theta_{23}=49.1^\circ$ \cite{Esteban:2024eli}, and $\mathring{\kappa}^{(6)}_{\mu \tau}=10^{-54}\,\mathrm{GeV}^{-2}$, we find using \cref{eq:2f} that the sharp peak should occur at $E\approx 1.6\times10^8\,\mathrm{GeV}$, which agrees very well with the numerical results in the three-flavor case shown in \cref{fig:flavor_fraction} (black line, right panel). What effectively occurs at this `peak' energy is the vanishing of the diagonal mixing matrix element $U_{\mu 2}$. This leads to strong attenuation of the muon neutrino component near the peak energy, resulting in the abrupt rise in the tau neutrino fraction. However, we also find that this peak is too narrow to have any significant impact on the tau neutrino event rate.

\section{Cosmogenic Neutrino Flux and \boldmath{$\nu_\tau$} Rates with LIV}
\label{sec:flux_det_count}
\noindent
In what follows, we discuss the techniques for calculating the cosmogenic neutrino flux (\cref{subsec:flux}) and then proceed to the calculation of the tau neutrino event rates in the presence of LIV at GRAND and POEMMA (\cref{subsec:tau_count}).


\begin{figure}[t]
    \centering
       \includegraphics[width=0.7\textwidth]{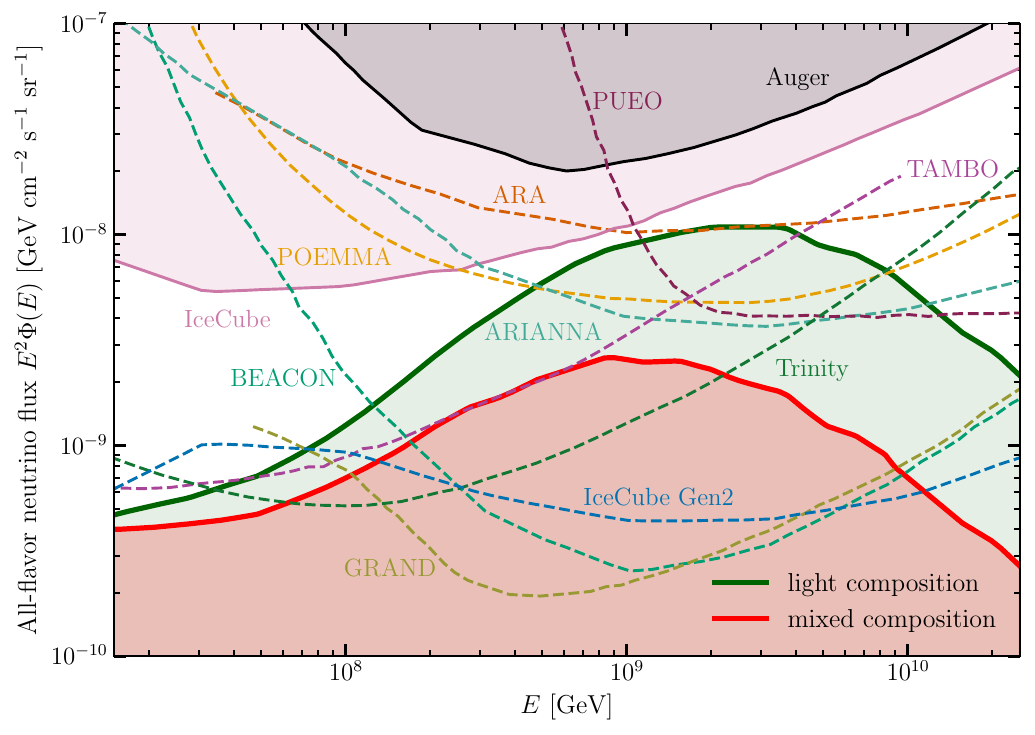}
       \caption{Cosmogenic neutrino fluxes considered in our analysis (red and green lines) are shown along with the current constraints from Auger \cite{PierreAuger:2019ens} and IceCube \cite{IceCubeCollaborationSS:2025jbi} and projected sensitivities for a number of experiments (POEMMA taken from Ref.~\cite{Cummings:2020ycz}, TAMBO from~\cite{Thompson:2025N8},  while the rest are adopted from Ref.~\cite{Ackermann:2022rqc}).}
    \label{fig:flux_sensitivity}
\end{figure}

\subsection{Cosmogenic Neutrino Flux}
\label{subsec:flux}
\noindent
UHECRs have been detected in various experiments, including the High Resolution Fly's Eye (HiRes)~\cite{HiRes:2009fiy}, the Telescope Array~\cite{TelescopeArray:2008toq}, and the Pierre Auger Observatory~\cite{PierreAuger:2015eyc}. 
They consist of protons and heavy nuclei with energies up to about $10^{21}\,\mathrm{eV}$. 
The relative contributions of protons and heavier nuclei are energy-dependent, and there is a known tension between the results reported by HiRes and Telescope Array and those from the Pierre Auger Observatory at the highest energies~\cite{HiRes:2009fiy,TelescopeArray:2018bep,Plotko:2022urd,PierreAuger:2023htc}. 
Among UHECRs, protons are the most efficient producers of cosmogenic neutrinos \cite{Hooper:2004jc, Ave:2004uj}, and therefore the relative composition of protons and heavier nuclei impacts the prospects for cosmogenic neutrino detection. \textcolor{black}{In this work, we compute the cosmogenic neutrino flux assuming a mixed mass composition of the UHECRs at the highest energies.} Our two representative benchmark fluxes (red and green lines in \cref{fig:flux_sensitivity}) are allowed in light of constraints from neutrino experiments, with the most stringent bound coming from IceCube's recent analysis of 12.6 years of data~\cite{IceCubeCollaborationSS:2025jbi} (pink region in \cref{fig:flux_sensitivity}). 
\textcolor{black}{The magnitudes of the fluxes calculated in astrophysical studies \cite{AlvesBatista:2018zui,vanVliet:2019nse,Ackermann:2022rqc,Globus:2025tud,Muzio:2025gbr,Das:2025vqd} and considered in new-physics studies \cite{Leal:2025eou,Machado:2025ltu} are in agreement with our benchmark flux shown in red. The other benchmark, shown in green, corresponds to a higher flux, likely constrained by the Pierre Auger measurements; however, given the aforementioned tension between the Pierre Auger and Telescope Array measurements, we consider this benchmark a viable alternative.}

In \cref{fig:flux_sensitivity}, we also show several projected sensitivities, including those for GRAND and POEMMA, experiments in the focus of our work.

\begin{table}[t]
    \centering
    \textcolor{black}{
    \begin{tabular}{|c|c|c|c|}
    \hline
    UHECR Parameter &
    Impact on Cosmogenic Neutrino Flux &
    \begin{tabular}{@{}c@{}}Changes\\Total Flux?\end{tabular} &
    \begin{tabular}{@{}c@{}}Changes\\$\nu_\alpha$ Flavor?\end{tabular} \\
    \hline
    \hline
    Spectral index $\gamma$ & Hard spectrum (small $\gamma$) can yield a high flux. & \checkmark & $\times$ \\
    \hline
    Mass composition & Presence of heavy nuclei results in suppressed & & \\
    & flux. The neutrino flavor composition at production & \checkmark & \checkmark \\
    & depends on the UHECR mass composition. & & \\
    \hline
    Redshift evolution $m$ & $\sim (1+z)^m$ & \checkmark & $\times$ \\
    \hline
    Emissivity & The neutrino flux scales with the emissivity factor. & \checkmark & $\times$ \\
    \hline
    \end{tabular}
    }
    \caption{\textcolor{black}{The model parameters to generate neutrino flux and their qualitative contribution.}}
    \label{tab:flux_features}
\end{table}

\textcolor{black}{The neutrino flux from UHECRs depends on multiple factors listed in the first column of ~\cref{tab:flux_features}. All these parameters affect the cosmic ray spectrum and subsequently the total neutrino flux. Neutrino production mechanism is dominated by pion decay for the proton-dominated UHECRs, while beta decay also contributes for the mixed-composition of UHECRs. Let us remind the reader that the flavor composition of neutrinos from photopion production and beta decay are $(f_e : f_\mu : f_\tau)_S = (1/3:2/3:0)$ and $(f_e : f_\mu : f_\tau)_S = (1:0:0)$, respectively. Thus, the neutrino flavor composition at Earth carries information on the UHECR mass composition. }

\textcolor{black}
{Let us now describe the procedure for computing the benchmark cosmogenic neutrino fluxes. 
We use \texttt{SimProp-v2r4}~\cite{Aloisio:2017iyh} to generate Monte Carlo (MC) events, where UHECRs, are injected uniformly at different redshifts (up to $z=10$) with injection energies ranging from $10^{17}~\rm eV$ to $10^{21}~\rm eV$. The mass composition of UHECRs is inferred from the depth of the shower maximum,~  $X_{\text{max}}$~\cite{Bl_mer_2009}. This quantity is the atmospheric depth at which the extensive air shower induced by a cosmic ray reaches its maximum development before beginning to attenuate. Light nuclei carry more energy per nucleon than heavy nuclei, allowing the resulting air showers to develop deeper in the atmosphere and yielding a larger average $X_{\mathrm{max}}$. Through this method, the UHECR mass composition has been reported in the range $[10^{16.5},10^{18.5}]~\rm eV$  and $[10^{18.5},10^{20}]~\rm eV$ by Telescope Array Low-energy Extension~\cite{TelescopeArray:2026rdu} and  Pierre Auger~\cite{PierreAuger:2024flk}, respectively. The former presented an increase of the fraction of heavy nuclei with energy; however, proton-dominated spectrum is preferred above $10^{17}\rm eV$. In contrast, the latter reported increasing fraction of heavy nuclei in UHECRs at the highest energies. Furthermore, Pierre Auger also reported best-fit parameters for the composition based on the measurements above $6\times 10^{17}~\rm eV$, separating low-energy and high-energy components for different model scenarios \cite{PierreAuger:2022atd}. While the reported best-fit composition is energy-bin dependent, in our analysis, for simplicity, the fraction of specific nuclei in the UHECR spectrum is fixed across energies. In general, we consider proton, nitrogen, silicon, and iron contribution to our model fluxes shown in \cref{fig:flux_sensitivity}. The two benchmark scenarios are constructed with different UHECR mass compositions: $f_{\text{H}} = 0.8,~f_{\text{N}}=0.1,~f_{\text{Si}}=0.1$ for the 'light composition' flux (green line in \cref{fig:flux_sensitivity}) and $f_{\text{H}} = 0.1,~f_{\text{N}}=0.3,f_{\text{Si}}=0.4,~f_{\text{Fe}}=0.2$ for the 'mixed composition' flux (red line).}

\textcolor{black}
{For the redshift evolution of the sources, we employ the star formation rate (SFR)~\cite{Yuksel:2008cu, Wang:2011qc}. The flux normalization is determined by source emissivity at $z=0$; see discussion in Ref.~\cite{Murase:2018utn}. For all the nuclei species, we take the source emissivity to be $ 10^{45}~\rm{erg/Mpc^3/yr}$ and the spectral index to be $\gamma_g=2$. We note that a single power law may not suffice to describe the cosmogenic neutrino flux as Pierre Auger reported hard spectrum above the ankle at 5~EeV~\cite{PierreAuger:2016use}.}

\textcolor{black}
{As nuclei propagate through the CMB and EBL, they produce secondary neutrinos via photopion production and radioactive beta decay, as described in \cref{sec:flavor_ratio_evolution}. 
The EBL model employed in our calculation is adopted from Refs.~\cite{Stecker:2005qs,Stecker:2006eh}. 
Using the framework described above together with \texttt{SimProp-v2r4}, one obtains the all-flavor cosmogenic neutrino flux shown in \cref{fig:flux_sensitivity}. One can infer from the figure that the benchmark flux computed with higher proton fraction yields a larger cosmogenic neutrino flux, as expected. }

\textcolor{black}{As we are interested in the tau neutrino flux at Earth, we interface \texttt{SimProp-v2r4} with the neutrino flavor transition probability framework in the presence of LIV by implementing \cref{eq:probb}; see also \cref{eq:ao}, which is employed for flavor evolution from production to detection sites. Importantly, the redshift $z$ at which a neutrino is produced, and which enters \cref{eq:probb}, does not correspond to the injection redshift of the nuclei for any given MC event. We extract the redshift corresponding to neutrino production using \texttt{SimProp-v2r4}, thereby ensuring a proper implementation of neutrino flavor transition probabilities. }

To summarize, each MC event leading to neutrino production in \texttt{SimProp-v2r4} has two redshifts recorded: nuclei injection redshift and neutrino production redshift. The cosmogenic neutrino fluxes, as presented in \cref{fig:flux_sensitivity}, are computed by integrating over the former, accounting for sources at different redshifts. This follows from Eq.~(B.4) and the subsequent description in Ref.~\cite{Aloisio:2017iyh}. For flavor transition computation, the neutrino production redshift is incorporated in \cref{eq:probb} and \cref{eq:ao}. We modify the integrand by including the LIV-induced flavor transition probability. Subsequently, the integral yields the LIV-dependent neutrino flux of a specific flavor; in our case, we focus on the tau flavor. For the purpose of reproducibility, our data files and the Python code are available on~\gitlink. Armed with the tau neutrino flux $\Phi_{\nu_\tau}(E)$, we now move to the calculation of tau neutrino event rates.

\subsection{Tau Neutrino Event Rates at GRAND and POEMMA}
\label{subsec:tau_count}
\noindent
Both Giant Radio Array for Neutrino Detection (GRAND)~\cite{GRAND:2018iaj} and Probe of Extreme Multi-Messenger Astrophysics (POEMMA)~\cite{Olinto:2017xbi, POEMMA:2020ykm} will primarily detect neutrinos via Earth-skimming tau neutrinos. Specifically, tau leptons are produced inside the Earth through charged-current interactions of cosmogenic tau neutrinos. Tau leptons can then traverse the Earth and emerge into the atmosphere, where they decay. This leads to the production of an extensive air shower. GRAND will be able to detect radio waves emitted from the air shower due to the geomagnetic~\cite{Kahn1966, Scholten:2007ky} and the Askaryan effect~\cite{Askaryan:1961pfb,Askaryan1965}. POEMMA, on the other hand, will not rely on radio signals, but instead will detect Cherenkov radiation from the air showers. Interestingly, POEMMA's precursor, POEMMA-Balloon with Radio (to be launched in 2027)~\cite{Adams:2026cpe}, will be sensitive to both radio and optical signals during its flights, although its sensitivity will be much weaker compared to POEMMA.

\subsubsection{GRAND}
\noindent
GRAND is a ground-based array of radio antennas designed to detect the radio emission from air showers. It is expected to have two arrays of 10,000 antennas each, deployed around 2030~\cite{Martineau-Huynh:2025hkr}. The end game is an array of 200,000 antennas, and our analysis is performed for such a configuration. 

The number of expected tau neutrino events reads 
\begin{align}
    N_{\nu_\tau}(\mathring{\kappa}^{(d)}_{\alpha \beta}) = T \Delta\Omega\int dE \, \Phi_{\nu_\tau} (E,\,\mathring{\kappa}^{(d)}_{\alpha \beta}) \,A_{\rm eff}(E) \,,
    \label{eq:NtauGRAND}
\end{align}
where $T$ is the data-taking time, $\Delta\Omega$ is the field of view, $\Phi_{\nu_\tau}(E,\,\mathring{\kappa}^{(d)}_{\alpha \beta})$ is the energy- and LIV-dependent tau neutrino flux (see \cref{subsec:flux} for calculation details), and $A_{\rm eff}$ is the direction-averaged effective area of the detector adopted from Ref.~\cite{GRAND:2018iaj} (Fig.~25). The neutrino energy integral is performed over an interval between 1~PeV and 100~EeV. The field of view for neutrinos is reported to be between $\theta = 85^\circ$ and $\theta = 95^\circ$~\cite{GRAND:2018iaj}, leading to a solid angle of observation $\Delta\Omega = 2\pi(\cos 85^\circ - \cos 95^\circ) \approx 1\,\mathrm{sr}$. We consider $T=10$ years for GRAND. 

\begin{figure}[t]
    \centering
    \includegraphics[width=0.52\linewidth]{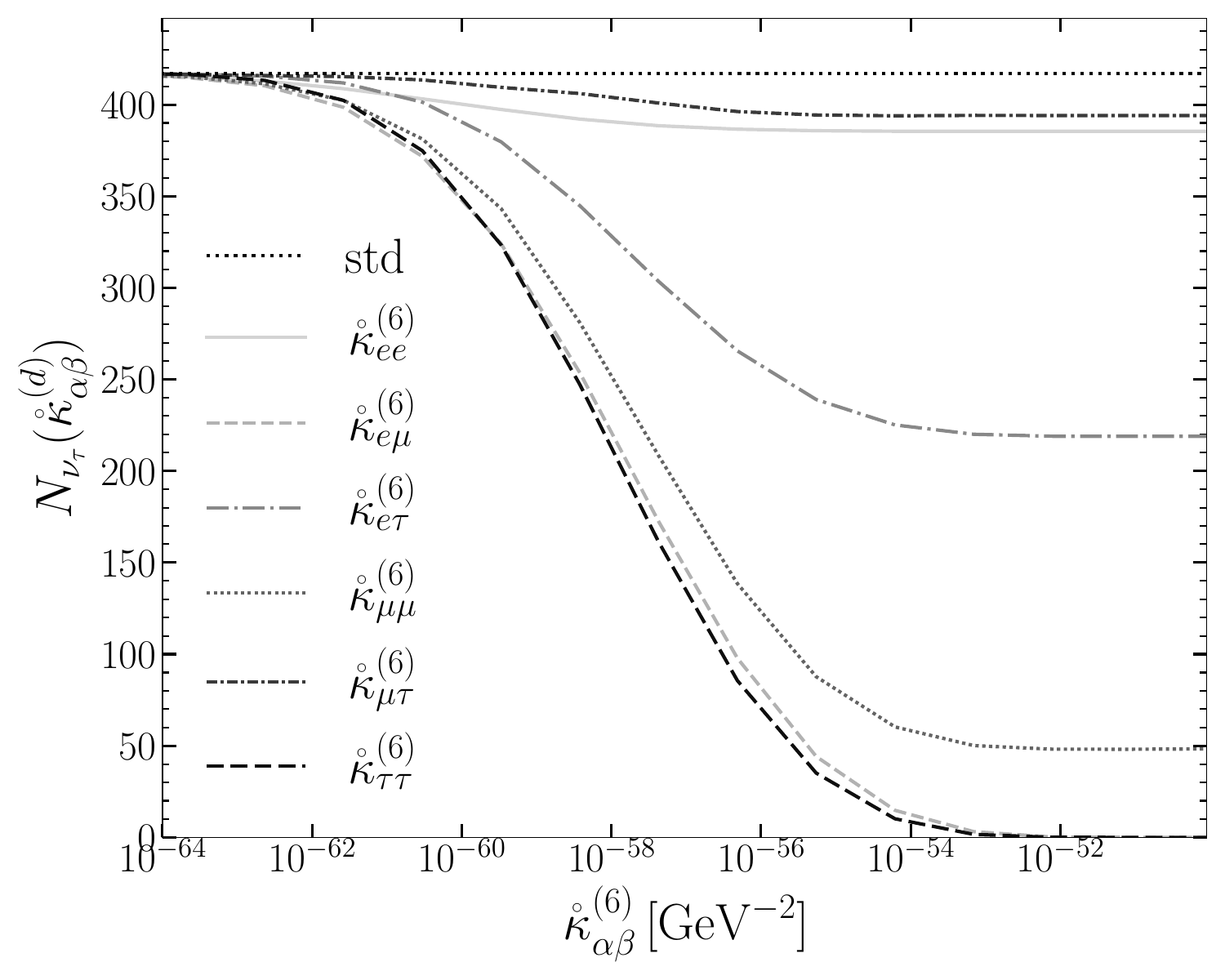}
    \caption{Number of expected tau neutrino events at GRAND with 10~years of data taking as a function of the LIV parameters corresponding to various flavor indices and for operator dimension $d=6$. Events are computed for the cosmogenic neutrino flux with light nuclei composition. The dotted line indicates the expected number of tau neutrino events in the absence of LIV.}
    \label{fig:tau_count}
\end{figure}

The number of tau-neutrino events at GRAND for six LIV parameters corresponding to different flavor indices at $d=6$ is presented in \cref{fig:tau_count}. The \textcolor{black}{dotted horizontal} line in the figure represents the expected number of tau neutrino events for standard neutrino oscillations in the absence of LIV, which reads $N_\text{obs}=\textcolor{black}{417}$ for the light composition ($N_\text{obs}=\textcolor{black}{102}$ for the mixed case). We generally observe that LIV effects decrease the number of expected tau neutrino events compared to the standard scenario. This change in the number of events is, however, relatively mild for $ee$ and $\mu\tau$ combinations of flavor indices. This can be explained using \cref{fig:flavor_fraction}, where we found that for these two cases, the change in flavor composition between scenarios with larger values of LIV and the no-LIV case is least pronounced.

The background events to the radio signal generated from ultra-high-energy tau neutrinos at GRAND could arise from stationary noise, transient sources, and UHECRs~\cite{GRAND:2018iaj}. The former two can be chiefly removed by implementing optimal choices of frequency range, timing profile, directionality, and polarization. Further, while there will be about a billion cosmic rays above 100~PeV detected at GRAND in a decade of data taking time, angular cut and shower maximum cut can reduce this background to $\sim 0.1$ per year \cite{GRAND:2018iaj}. Given the above, we consider background-free realization in our analysis.

\begin{figure}[t]
    \centering
    \includegraphics[width=0.52\linewidth]{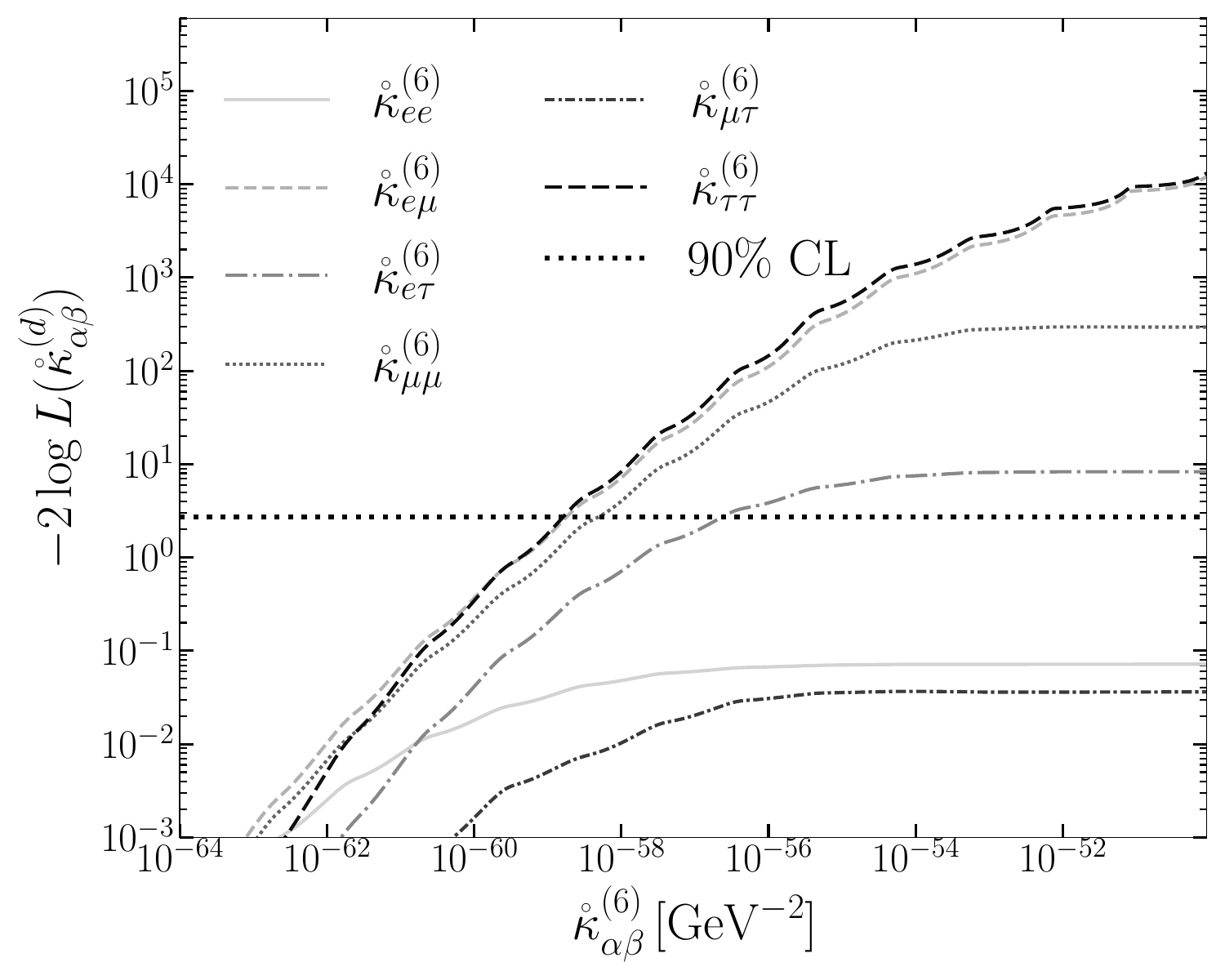}
    \caption{The log-likelihood, $-2 \log L$, associated with tau neutrino observations at GRAND is shown as a function of the LIV parameter $\mathring{\kappa}^{(6)}_{\alpha \beta}$. It is constructed from the expected event counts shown in \cref{fig:tau_count}. The dotted horizontal line shows the nominal value of $-2 \log L$ at which we determine the 90\% CL projected sensitivities for one degree of freedom.}
    \label{fig:tau_chi2}
\end{figure}

\subsubsection{POEMMA}
\noindent 
POEMMA will consist of a pair of space-based satellites sensitive to air fluorescence and optical Cherenkov radiation, and is expected to be launched in the 2030s.

For POEMMA, the number of expected tau neutrino events reads 
\begin{align}
    N_{\nu_\tau}(\mathring{\kappa}^{(d)}_{\alpha \beta}) = T \int dE \, \Phi_{\nu_\tau} (E,\,\mathring{\kappa}^{(d)}_{\alpha \beta}) \,\langle A\Omega \rangle (E) \,.
    \label{eq:NtauPOEMMA}
\end{align}
The only qualitative difference compared to \cref{eq:NtauGRAND} is the replacement of the effective area with the geometric aperture $\langle A\Omega \rangle$, which accounts for the solid angle of detection of the Cherenkov emission. The energy integration interval is the same as for GRAND. The geometric aperture is taken from Fig.~22 of Ref.~\cite{Cummings:2020ycz}. Their analysis considers detection prospects for different minimum Cherenkov photon densities $\rho_{\rm thr}$, defined as the threshold photon density required to trigger a signal over the background. The aperture we use corresponds to the nominal, conservative threshold of $\rho_{\rm thr}=20~\text{photons}/\text{m}^2$. POEMMA is designed for a nominal operation period of $T=5$ years~\cite{POEMMA:2020ykm}. We include an additional factor of $0.2$, assuming a 20\% duty cycle, which then yields an effective data-taking period of one year.

\textcolor{black}{For POEMMA, the expected number of tau neutrinos in the absence of LIV is $N_\text{obs}\simeq \textcolor{black}{8}$ for the benchmark cosmogenic neutrino flux with high-proton content. \textcolor{black}{This may be insufficient to put constraint on new physics.} Further improvements can be possible with reducing the detector threshold, increasing the number of employed satellites, and with longer flight time or larger duty cycle. In this work, we consider an enhancement of 5 to the current sensitivity reach, and we will be \textcolor{black}{referring} to such configuration as 'POEMMAx5' in what follows.  We also assume that background events to Cherenkov signals from tau neutrinos at POEMMA are negligible~\cite{Venters:2019xwi}. }

\section{LIV Sensitivity at GRAND and POEMMA}
\label{sec:exp_sens}
\noindent
Having established the framework for calculating the number of tau neutrino events at GRAND and POEMMA, we define the log-likelihood 
\begin{equation}
  -2 \log L =  2 \left( N_{\text{exp}} - N_{\text{obs}} + N_{\text{obs}} \log \frac{N_{\text{obs}}}{N_{\text{exp}}} \right)+\frac{\xi^2}{\sigma_\xi^2}\,.
    \label{eq:chi2}
\end{equation}
\textcolor{black}{Here, $N_{\text{exp}}= (1+\xi) N_{\nu_\tau}(\mathring{\kappa}^{(d)}_{\alpha \beta})$ is the expected number of tau neutrino events in the presence of LIV, while $N_{\text{obs}}=N_{\nu_\tau}(\mathring{\kappa}^{(d)}_{\alpha \beta}=0)$ is the expected number of measured events assuming the absence of any LIV effects. $\xi$ is the nuisance parameter for the flux normalization. We profile over $\xi$ with a mean value of $0$ and $\sigma_\xi=0.3$ (uncertainty of 30\%) as justified in what follows. From  \cref{fig:flux_sensitivity}, by comparing benchmark fluxes with sensitivities, we infer that IceCube Gen-2 radio is expected to detect $\sim{10}$ and $\sim{40}$ neutrino events for the light and mixed composition fluxes, respectively. With statistical uncertainty of $1/\sqrt{N}$, we obtain 30\% flux uncertainty as a conservative estimate in our calculation.} Note that for  GRAND and POEMMA, we quoted $N_{\text{obs}}$ for both mass composition scenarios in \cref{subsec:tau_count}.

\begin{figure}[hb]
    \centering
       \includegraphics[width=0.8\textwidth]{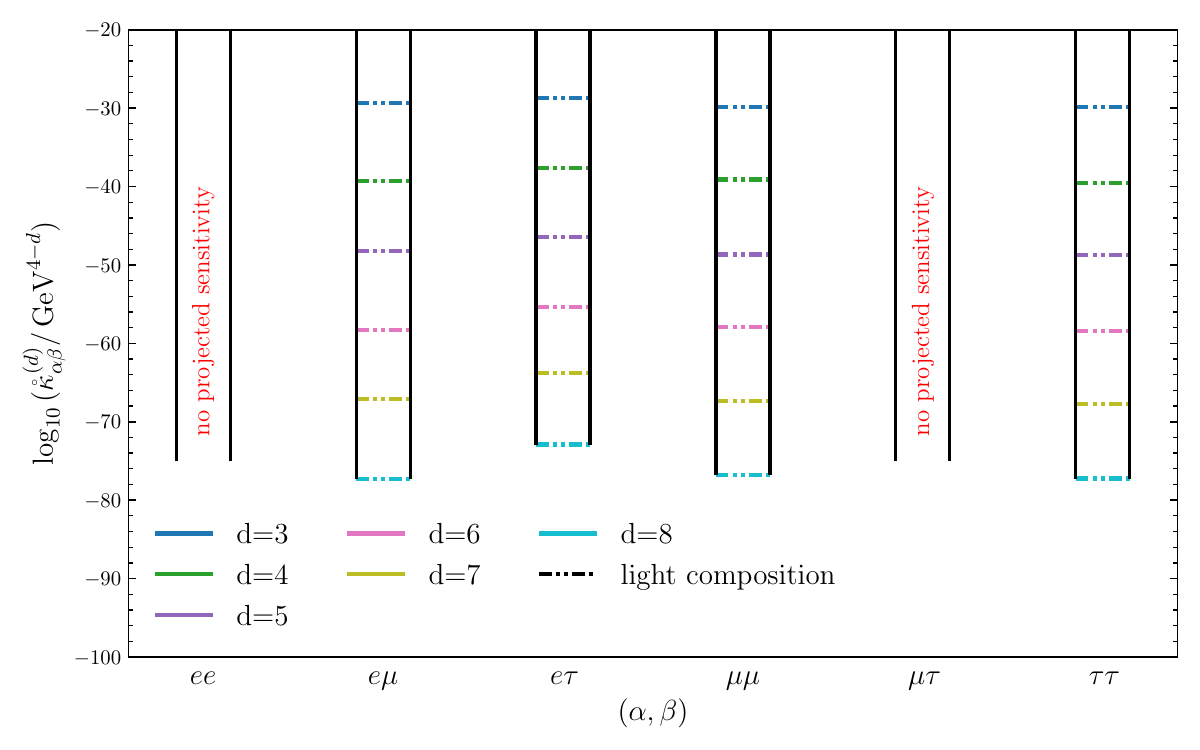}
    \caption{Projected sensitivities to LIV parameters for the POEMMAx5 experiment, assuming that only a single LIV parameter $\mathring{\kappa}^{(d)}_{\alpha \beta}$ is nonzero. Operator dimensions $d=3$ to $8$ are considered for all six independent combinations of flavor indices. \textcolor{black}{Sensitivities are obtained using the cosmogenic neutrino flux calculated for light mass UHECR composition.}}
    \label{fig:param_sens_poemma}
\end{figure}

In \cref{fig:tau_chi2}, for GRAND, we show $-2 \log L$ as a function of $\mathring{\kappa}^{(6)}_{\alpha \beta}$ for different flavor indices $\{\alpha,\beta\}$. We observe from the figure that the log-likelihood at which we set the 90\% CL projected sensitivities is reached for all parameters except the $\mu\tau$ and $ee$ case. This is expected, as for these two cases we found in \cref{fig:tau_count} the least pronounced LIV-induced change in tau neutrino event counts. Based on this, we expect the projected sensitivities for $\mathring{\kappa}^{(d)}_{ee}$ and $\mathring{\kappa}^{(d)}_{\mu\tau}$ to be significantly weaker compared to the other parameters, or even non-existent.

\textcolor{black}{
The 90\% CL projected sensitivities for LIV operators with dimensions $d=3$ through $d=8$ and for all six combinations of flavor indices are presented for POEMMAx5 in \cref{fig:param_sens_poemma}. Importantly, in calculating these sensitivities, we assumed that only a single LIV parameter $\mathring{\kappa}^{(d)}_{\alpha \beta}$ is nonzero. Our results correspond to the cosmogenic neutrino benchmark flux with light UHECR composition, since POEMMA will not be sensitive to the mixed composition one (see again \cref{fig:flux_sensitivity}). Even for the benchmark flux with a light UHECR composition, we find that POEMMAx5 yields \textcolor{black}{no sensitivity to LIV parameters with $ee$ and $\mu\tau$  flavor indices.}
}

When comparing results for all other flavor combinations across different dimensions, we observe that the sensitivity strongly improves with increasing $d$. This trend arises primarily from the energy dependence associated with LIV; see \cref{eq:kappa_compare_LIV_vac}, which illustrates how smaller and smaller parameters become testable as energy increases, with the scaling $E^{2-d}$.


\begin{figure}[ht!]
    \centering
       \includegraphics[width=0.8\textwidth]{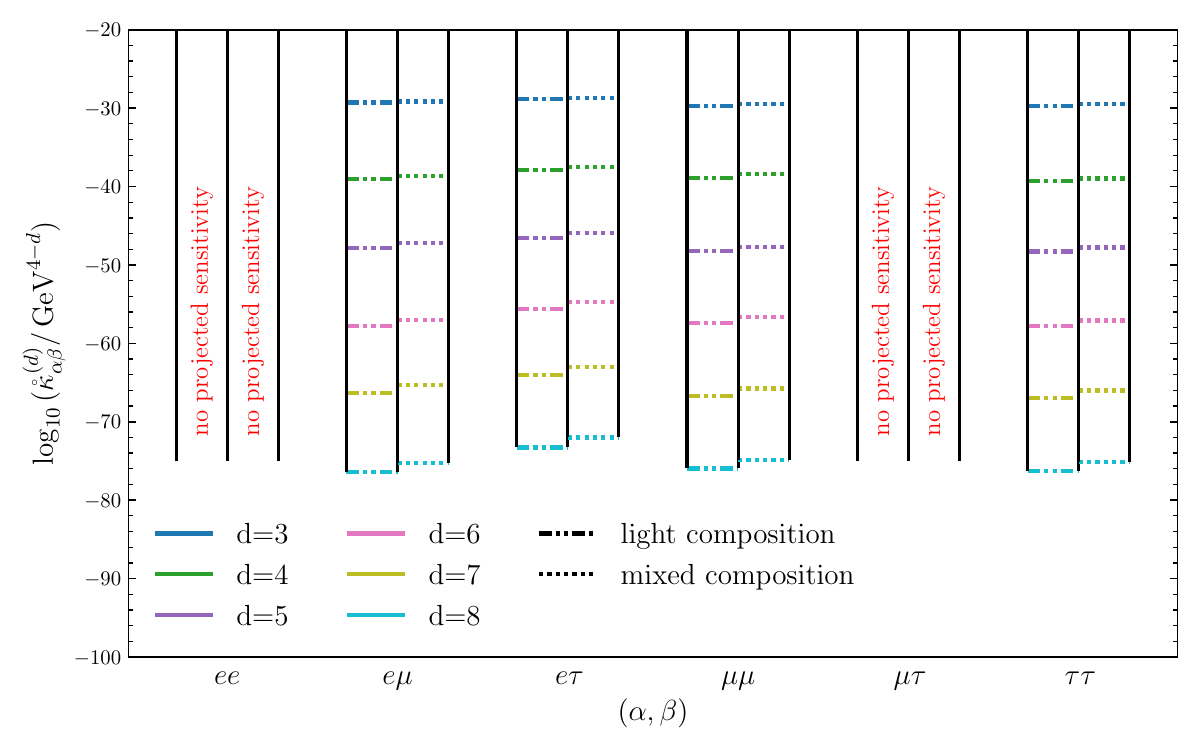}
    \caption{Projected sensitivities to LIV parameters for the GRAND experiment, assuming that only a single LIV parameter $\mathring{\kappa}^{(d)}_{\alpha \beta}$ is nonzero. Operator dimensions $d=3$ to $8$ are considered for all six independent combinations of flavor indices. \textcolor{black}{Sensitivities are obtained using the cosmogenic neutrino flux calculated for both light and mixed mass composition of UHECRs.}}
    \label{fig:param_sens_grand}
\end{figure}

In \cref{fig:param_sens_grand}, we show 90\% CL projected sensitivities at GRAND for LIV operators with dimensions $d=3$ through $d=8$ and for all six combinations of flavor indices. As in \cref{fig:param_sens_poemma}, in calculating these results, we assumed that only a single LIV parameter $\mathring{\kappa}^{(d)}_{\alpha \beta}$ is nonzero. 

\textcolor{black}{For each pair of flavor indices, we present sensitivities corresponding to the cosmogenic neutrino flux calculation using light (dot-dashed) and mixed (dotted) mass compositions of the UHECRs. For the former case, the sensitivity is generally stronger. This is because, for a given emissivity, the cosmogenic neutrino flux is higher in the light composition scenario, both in the cases with and without LIV, and the log-likelihood grows with the number of tau neutrino events. In accord with the results presented in \cref{fig:tau_chi2} and \cref{fig:param_sens_poemma}, we find no sensitivity for the $ee$ and $\mu\tau$ coefficients for both flux scenarios and all considered $d$. }

\textcolor{black}{For precise values of projected sensitivities (colored horizontal lines in \cref{fig:param_sens_poemma,fig:param_sens_grand}) for both GRAND and POEMMAx5, we refer the reader to \cref{tab:results} in appendix~\ref{app:supp}. GRAND and POEMMAx5 exhibit similar projected sensitivities on LIV parameters, for the light composition flux and for all considered dimensions of LIV operators, even though GRAND has a visibly stronger UHE neutrino sensitivity (\cref{fig:flux_sensitivity}). This can be explained through the overlap between the neutrino flux and the detector sensitivity. POEMMA, although its flux sensitivity is weaker than that of GRAND, will be able to detect neutrinos from the peak of the light composition flux, from $\sim 5\times 10^8~\rm GeV$ to $\sim 10^{10}~\rm GeV$~(\cref{fig:flux_sensitivity}). Consequently, the predicted event rates show a stronger $\mathring{\kappa}^{(d)}_{\alpha\beta}$-dependence, resulting in a LIV parameter sensitivity comparable to that of GRAND. Note that, unlike POEMMAx5, GRAND is sensitive to $\mathring{\kappa}^{(d)}_{e\tau}$. However, neither of the considered experiments is sensitive to the parameters $\mathring{\kappa}^{(d)}_{ee}$ and $\mathring{\kappa}^{(d)}_{\mu\tau}$ at 90\% CL.} 

For all combinations of flavor indices apart from $ee$ and $\mu\tau$, the derived sensitivities can very roughly be expressed as $\mathring{\kappa}^{(d)}_{\alpha\beta}\sim 10^{-10\times d}~\text{GeV}^{4-d}$. For both experiments, the sensitivities exceed existing constraints from IceCube using PeV neutrinos \cite{IceCube:2021tdn} by orders of magnitude. This improvement originates from the polynomial dependence of the LIV parameters on the neutrino energy, which leads to a significantly larger effect for cosmogenic neutrinos compared to current IceCube datasets. We note that the GRAND sensitivities for a single nonzero LIV parameter were also derived in Ref.~\cite{Testagrossa:2023ukh} for four different combinations of flavor indices. We find comparable results, despite different assumptions on \textcolor{black}{the origin of ultra-high-energy} neutrino fluxes.

Quantum gravity effects at the Planck scale are a well-motivated source for Lorentz invariance violation. The dimension of $\mathring{\kappa}^{(d)}_{\alpha \beta}$ (see \cref{eq:kappa_compare_LIV_vac}) allows us to compute the expected magnitude of these parameters, assuming they arise from LIV near the Planck scale. This leads to
$\mathring{\kappa}^{(d)}_{\alpha \beta}\sim 1/m_\text{Pl}^{d-4}$, where $m_{\text{Pl}}=1.22\times 10^{19}$ GeV. Cases with at least one power of the Planck mass in the denominator are of interest; for $d=(5,6,7,8)$ we find {$\mathring{\kappa}^{(d)}_{\alpha \beta}\sim (8.2\times10^{-20} \, \text{GeV}^{-1},\, 6.7\times10^{-39} \, \text{GeV}^{-2}, \, 5.5\times 10^{-58}\, \text{GeV}^{-3},\, 4.5\times 10^{-77}\, \text{GeV}^{-4})$}, which are all within the sensitivity reach of GRAND and POEMMAx5. This implies that the quantum-gravity-motivated parameter space is testable with the upcoming neutrino experiments.

\begin{figure}[t]
    \centering
       \includegraphics[width=0.65\textwidth]{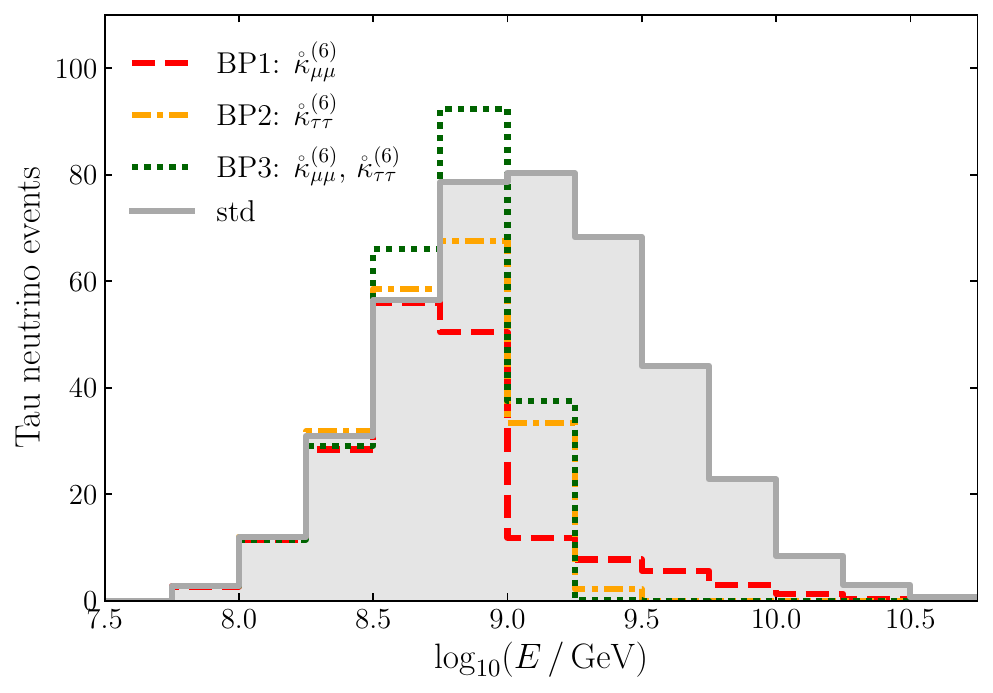}
    \caption{We present the number of expected tau neutrino events at GRAND in energy bins between $E=10^{7.5}$ GeV and $E=10^{11}$ GeV for three benchmark scenarios (BP1, BP2, BP3) with nonzero LIV, as well as the no-LIV case with standard flavor transition (gray). A cosmogenic flux assuming light composition is employed. BP1 (red) corresponds to 
    $\mathring{\kappa}^{(6)}_{\mu\mu} = 1\times 10^{-57}~\mathrm{GeV}^{-2}$, BP2 (orange) corresponds to $\mathring{\kappa}^{(6)}_{\tau\tau} = \textcolor{black}{1}\times 10^{-58}~\mathrm{GeV}^{-2}$, and BP3 (green) includes both of these parameters simultaneously, each with the same value as in BP1 and BP2 scenarios. We find that BP1 and BP2 will be strongly excluded with future GRAND data, while BP3, with both LIV parameters active simultaneously, does not yield a comparably significant deviation in event counts from the no-LIV case and will not be testable at GRAND.}
    \label{fig:param2d_blindspot}
\end{figure}

Above results, summarized in \cref{fig:param_sens_poemma,fig:param_sens_grand}, are obtained assuming only a single LIV parameter $\mathring{\kappa}^{(d)}_{\alpha \beta}$ to be nonzero. These sensitivities do not apply to the scenario featuring multiple nonzero LIV parameters. Let us illustrate this in \cref{fig:param2d_blindspot}, where we present the number of expected tau neutrino events in 10 years, recorded in energy bins between $E=10^{7.5}$ GeV and $E=10^{11}$ GeV at GRAND, assuming SFR source evolution for three benchmark points (BPs). 
BP1 (red) corresponds to a single nonzero LIV parameter $\mathring{\kappa}^{(6)}_{\mu\mu} = 1\times 10^{-57}~\mathrm{GeV}^{-2}$, while BP2 (orange) corresponds to a single nonzero parameter $\mathring{\kappa}^{(6)}_{\tau\tau} = \textcolor{black}{1}\times 10^{-58}~\mathrm{GeV}^{-2}$. 
Both parameters are chosen such that, when considered individually (with all other LIV parameters set to zero), they lie well within the projected sensitivity limit, i.e., will be constrained with GRAND data. Indeed, one can infer from \cref{fig:param2d_blindspot} the clear difference between red and orange histograms compared to the standard flavor transition with no LIV (gray). In the BP3 case (green), we take the above two parameters to be present simultaneously with the same values as in BP1 and BP2 scenarios. As can be seen from the figure, their interplay yields a total tau neutrino event count that is closer to the no-LIV case, and it turns out that such a combination of LIV parameter values cannot be constrained with future data from GRAND.

\begin{figure}[t]
    \centering
       \includegraphics[width=0.85\textwidth]{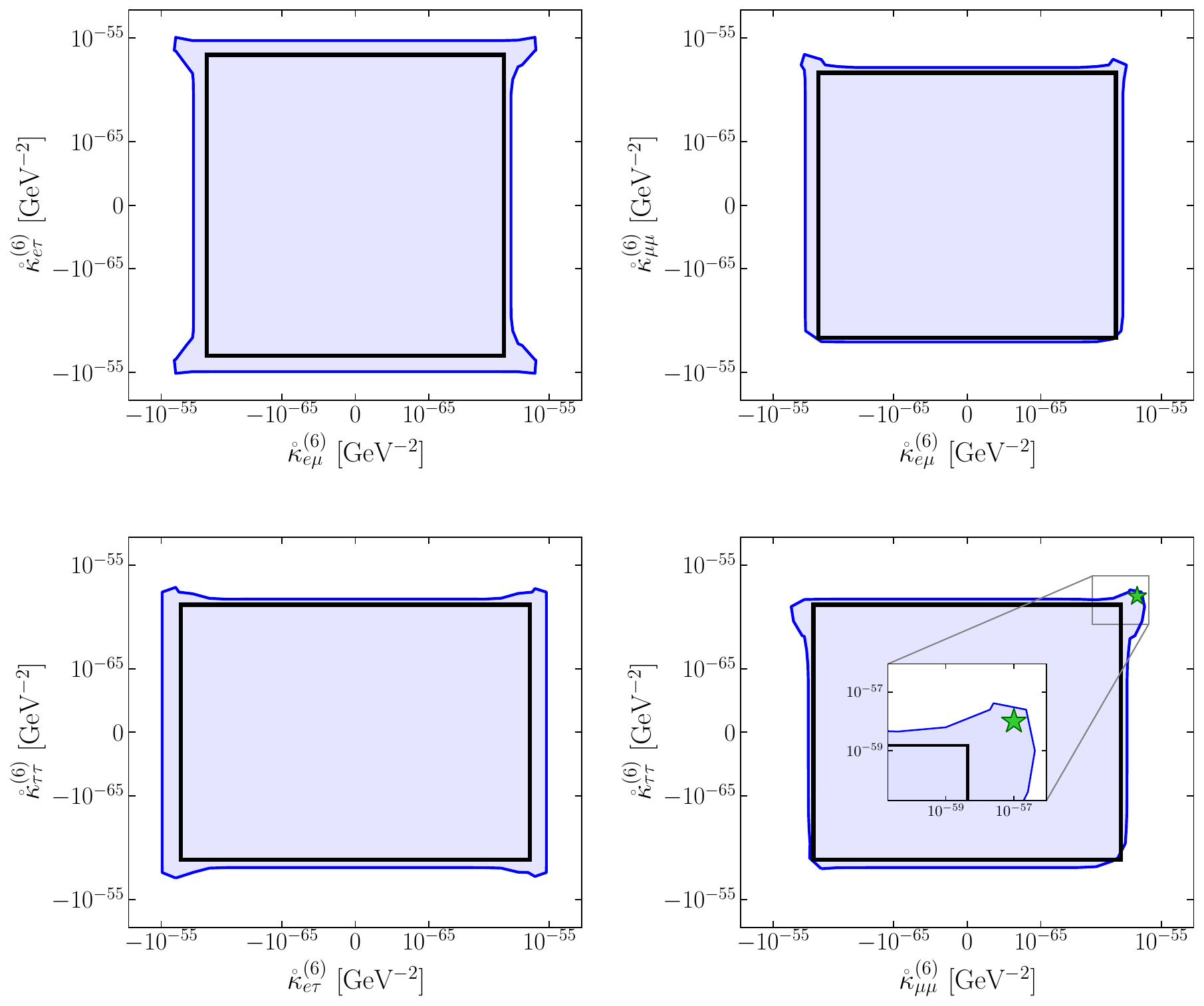}
    \caption{Each panel shows the projected sensitivity to a pair of LIV parameters, corresponding to $d=6$, for the GRAND experiment, computed using the cosmogenic neutrino flux with light composition (blue region). For comparison, we also show the sensitivities obtained when each parameter is considered individually (black lines).}
    \label{fig:param2d_selected}
\end{figure}

In \cref{fig:param2d_selected}, we present two-parameter sensitivities at GRAND for four pairs of LIV parameters corresponding to $d=6$ and various flavor indices. In each panel, the white region denotes the parameter range that can be probed with future GRAND data, while the blue region cannot. We also show black rectangles whose sides correspond to the sensitivities obtained in a single-parameter analysis presented in \cref{fig:param_sens_grand}, and tabulated in \cref{tab:results}. Across all panels, we observe a clear difference between the blue region and the region enclosed by black lines; they differ by as much as a few orders of magnitude in $\mathring{\kappa}^{(6)}_{\alpha\beta}$. The green star in the lower right panel corresponds to BP3 in \cref{fig:param2d_blindspot} and, as expected, lies outside the black square (can be probed with GRAND data in a single-parameter scenario) but inside the blue region (cannot be probed in a two-parameter scenario). The results in \cref{fig:param2d_blindspot,fig:param2d_selected} clearly imply that a naive single-parameter analysis cannot be directly applied to multi-parameter cases, where the interplay between different LIV parameters can nullify the effects of individual parameters when all others are set to zero. Given six $\mathring{\kappa}^{(d)}_{\alpha \beta}$ parameters for a fixed $d$, there are $15$ pairs of parameters in total. \textcolor{black}{Several combinations involving LIV parameters with $ee$ and/or $\mu\tau$ flavor indices yield no sensitivity; the projected sensitivities for the remaining combinations are shown in \cref{fig:param2d_remaining} in appendix~\ref{app:supp}.}


All presented projected sensitivities to LIV parameters are calculated for two benchmark cosmogenic neutrino fluxes shown in \cref{fig:flux_sensitivity} (green and red lines represent cases corresponding to light and mixed UHECR composition, respectively). While these are representative fluxes consistent with all present constraints, the true cosmogenic neutrino flux remains unknown. This flux can realistically be measured using the upcoming IceCube-Gen2 radio experiment \cite{IceCube-Gen2:2020qha} (compare its sensitivity shown in \cref{fig:flux_sensitivity} with our benchmark fluxes), which will be sensitive to all neutrino flavors. If such a flux is measured, the pipeline presented in this work can then be used to constrain LIV parameters, either in a single- or multi-parameter scenario. If no flux is observed, this would imply that UHECRs are heavy-nuclei dominated. If GRAND observes tau neutrino deficit, a characteristic of LIV (see \cref{fig:tau_count}), the measurement at IceCube-Gen2 radio will be crucial to determine if this deficit is coming from LIV or, instead, UHECR mass composition is dominated by heavy-nuclei. We therefore stress the necessity of having at least two detectors capable of measuring \textcolor{black}{ultra-high-energy} neutrinos; for a related discussion, see Ref.~\cite{Testagrossa:2023ukh}.

\section{Conclusions}
\label{sec:conclusions}
\noindent
In this work, we have explored the potential of upcoming experiments GRAND and POEMMA to probe LIV in the neutrino sector. This is achievable through the detection of cosmogenic neutrinos, whose flavor transition probabilities change in the presence of LIV operators.
We focused on tau neutrino events, expected to be identified in an essentially background-free setup where Earth-skimming tau neutrinos produce tau leptons that exit the Earth and decay,
generating detectable radio or optical signals. The difference in the expected number of tau neutrino events in the presence of LIV with respect to the standard case forms the basis for calculating sensitivities to the LIV parameters.
This allowed us to obtain the sensitivities to isotropic LIV operators of various dimensions, namely from $d=3$ to $d=8$; see \cref{eq:HLIV_kappa} for the form of these LIV terms. The derived sensitivities for the scenarios with a single nonzero LIV parameter are shown in \cref{fig:param_sens_poemma,fig:param_sens_grand,tab:results}. We generally find them to exceed constraints from lower-energy probes by dozens of orders of magnitude.

The strength of these results arises from the polynomial energy dependence ($E^{2-d}$) of the LIV terms (see \cref{eq:kappa_compare_LIV_vac}), which becomes particularly pronounced for ultra-high-energy cosmogenic neutrinos whose energies exceed those detected by IceCube by several orders of magnitude.
The improved reach of GRAND and POEMMA \textcolor{black}{(with the POEMMAx5 configuration, corresponding to a fivefold improvement in sensitivity, considered in this work)} is particularly apparent for higher-dimensional operators;
e.g., for $d=8$, the projected sensitivities to LIV parameters are about $30$ orders of magnitude stronger than the present constraints from IceCube. We also explored the sensitivities for pairs of LIV parameters; these are shown in \cref{fig:param2d_selected,fig:param2d_remaining}, and the differences compared to the single-parameter results imply that the single-parameter approach cannot be directly applied to multi-parameter cases.

In summary, the discovery of cosmogenic neutrinos, and more specifically, the detection of extreme-energy tau neutrinos, will lead either to drastic improvements in the LIV constraints or to the discovery of LIV  with far-reaching implications for our understanding of fundamental physics. 

\section{Acknowledgements}
\noindent
SRM gratefully acknowledges helpful discussions with Austin L. Cummings and Felix Schlüter. SRM would also like to thank Writasree Maitra for providing the flux sensitivities in ~\cref{fig:flux_sensitivity}. The work of VB and SRM is supported by the United States Department of Energy Grant No. DE-SC0025477. 


\bibliographystyle{JHEP}
\bibliography{refs}

\appendix

\newpage
\section{Supplementary Figures and Tables}
\label{app:supp}
\begin{figure}[h!]
    \centering
       \includegraphics[width=0.97\textwidth]{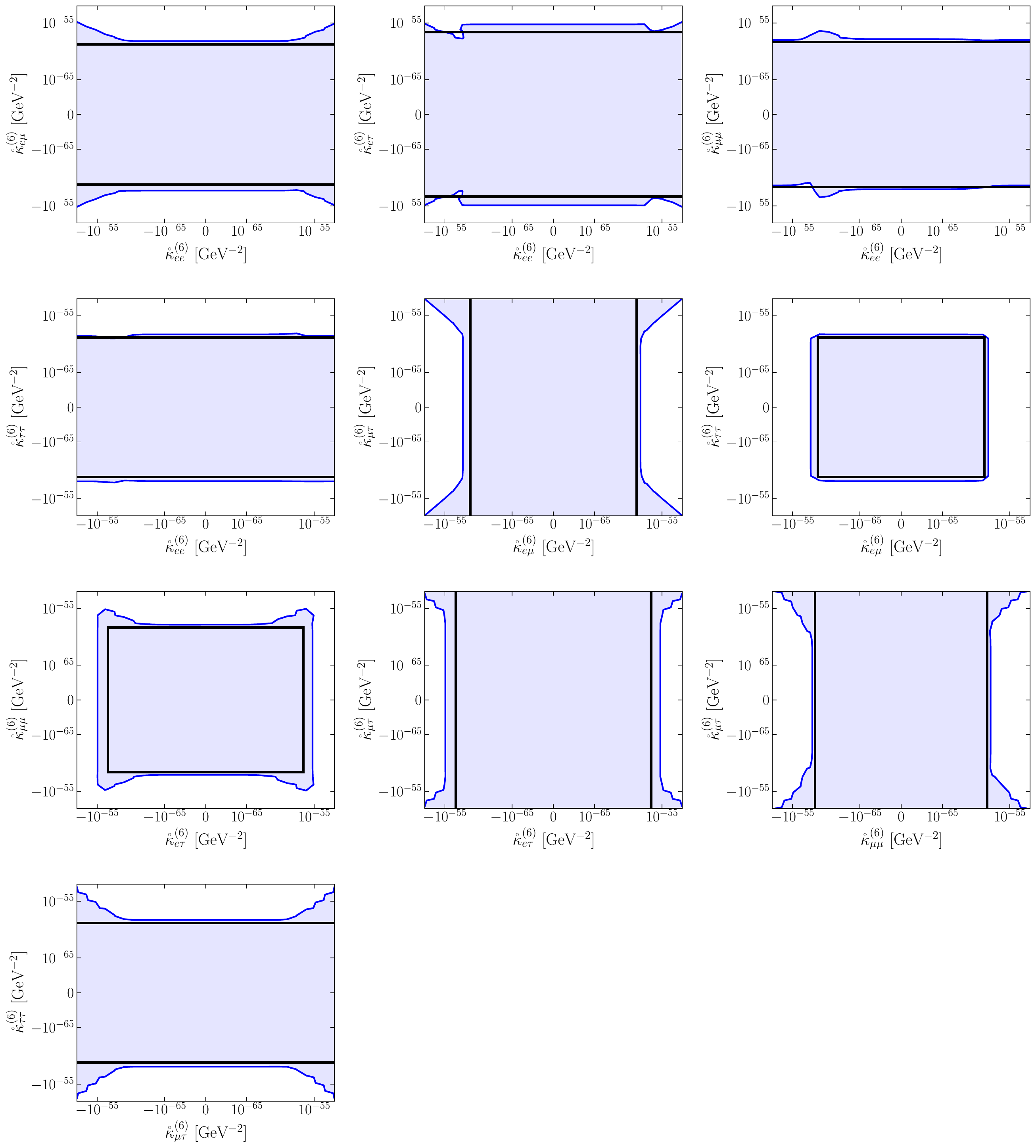}
    \caption{Each panel shows the projected sensitivity to a pair of LIV parameters, corresponding to $d=6$, for the GRAND experiment, computed using the cosmogenic neutrino flux with light composition (blue region). For comparison, we also show the sensitivities obtained when each parameter is considered individually (black lines).}
    \label{fig:param2d_remaining}
\end{figure}

\begin{table}[ht]
\centering
\renewcommand{\arraystretch}{1}
\setlength{\tabcolsep}{6pt}
\begin{tabular}{cc cc c}
\hline\hline
& & \multicolumn{2}{c}{\textsc{GRAND}} & \multicolumn{1}{c}{\textsc{POEMMAx5}} \\
\cmidrule(lr){3-4} \cmidrule(lr){5-5}
$d$ & Parameter & mixed composition & light composition & light composition \\
\hline
\multirow{6}{*}{3}
& $\mathring{\kappa}^{(3)}_{e\mu}$       & $7.1\times10^{-31}$ & $5.1\times10^{-31}$ & $4.0\times10^{-31}$ \\
& $\mathring{\kappa}^{(3)}_{e\tau}$      & $2.2\times10^{-30}$ & $1.5\times10^{-30}$ & $2.1\times10^{-30}$ \\
& $\mathring{\kappa}^{(3)}_{\mu\mu}$     & $3.0\times10^{-31}$ & $1.7\times10^{-31}$ & $1.3\times10^{-31}$ \\
& $\mathring{\kappa}^{(3)}_{\tau\tau}$   & $3.4\times10^{-31}$ & $1.8\times10^{-31}$ & $1.3\times10^{-31}$ \\
\hline
\multirow{6}{*}{4}
& $\mathring{\kappa}^{(4)}_{e\mu}$       & $2.0\times10^{-40}$ & $8.3\times10^{-41}$ & $5.3\times10^{-41}$ \\
& $\mathring{\kappa}^{(4)}_{e\tau}$      & $3.4\times10^{-39}$ & $1.4\times10^{-39}$ & $2.2\times10^{-39}$ \\
& $\mathring{\kappa}^{(4)}_{\mu\mu}$     & $3.6\times10^{-40}$ & $1.4\times10^{-40}$ & $8.0\times10^{-41}$ \\
& $\mathring{\kappa}^{(4)}_{\tau\tau}$   & $1.1\times10^{-40}$ & $5.1\times10^{-41}$ & $2.9\times10^{-41}$ \\
\hline
\multirow{6}{*}{5}
& $\mathring{\kappa}^{(5)}_{e\mu}$       & $5.9\times10^{-49}$ & $1.5\times10^{-49}$ & $5.5\times10^{-50}$ \\
& $\mathring{\kappa}^{(5)}_{e\tau}$      & $1.3\times10^{-47}$ & $2.8\times10^{-48}$ & $4.0\times10^{-48}$ \\
& $\mathring{\kappa}^{(5)}_{\mu\mu}$     & $1.9\times10^{-49}$ & $5.8\times10^{-50}$ & $2.2\times10^{-50}$ \\
& $\mathring{\kappa}^{(5)}_{\tau\tau}$   & $1.7\times10^{-49}$ & $5.1\times10^{-50}$ & $1.7\times10^{-50}$ \\
\hline
\multirow{6}{*}{6}
& $\mathring{\kappa}^{(6)}_{e\mu}$       & $9.4\times10^{-59}$ & $1.7\times10^{-59}$ & $5.1\times10^{-60}$ \\
& $\mathring{\kappa}^{(6)}_{e\tau}$      & $1.8\times10^{-56}$ & $2.4\times10^{-57}$ & $4.2\times10^{-57}$ \\
& $\mathring{\kappa}^{(6)}_{\mu\mu}$     & $2.3\times10^{-58}$ & $4.4\times10^{-59}$ & $1.2\times10^{-59}$ \\
& $\mathring{\kappa}^{(6)}_{\tau\tau}$   & $8.1\times10^{-59}$ & $1.5\times10^{-59}$ & $3.8\times10^{-60}$ \\
\hline
\multirow{6}{*}{7}
& $\mathring{\kappa}^{(7)}_{e\mu}$       & $4.3\times10^{-67}$ & $5.0\times10^{-68}$ & $7.8\times10^{-69}$ \\
& $\mathring{\kappa}^{(7)}_{e\tau}$      & $9.5\times10^{-65}$ & $9.0\times10^{-66}$ & $1.7\times10^{-65}$ \\
& $\mathring{\kappa}^{(7)}_{\mu\mu}$     & $1.7\times10^{-67}$ & $2.1\times10^{-68}$ & $4.0\times10^{-69}$ \\
& $\mathring{\kappa}^{(7)}_{\tau\tau}$   & $9.7\times10^{-68}$ & $1.2\times10^{-68}$ & $2.1\times10^{-69}$ \\
\hline
\multirow{6}{*}{8}
& $\mathring{\kappa}^{(8)}_{e\mu}$       & $5.5\times10^{-77}$ & $4.3\times10^{-78}$ & $5.6\times10^{-79}$ \\
& $\mathring{\kappa}^{(8)}_{e\tau}$      & $9.8\times10^{-74}$ & $5.2\times10^{-75}$ & $1.2\times10^{-74}$ \\
& $\mathring{\kappa}^{(8)}_{\mu\mu}$     & $1.4\times10^{-76}$ & $1.1\times10^{-77}$ & $1.7\times10^{-78}$ \\
& $\mathring{\kappa}^{(8)}_{\tau\tau}$   & $6.8\times10^{-77}$ & $5.4\times10^{-78}$ & $5.7\times10^{-79}$ \\
\hline\hline
\end{tabular}
\caption{90\% C.L. projected sensitivities on LIV parameters $\mathring{\kappa}^{(d)}_{\alpha\beta}$ for dimensions $d=3$ to $d=8$, derived for two flux models with light and mixed UHECR mass composition.}
\label{tab:results}
\end{table}


\end{document}